\documentclass[manuscript]{aastex}

\shorttitle{X-ray emission from the double
hot spot of 3C 351}
\shortauthors{Brunetti et al.}

\begin{document}


\title{{\it CHANDRA} DETECTION OF THE RADIO AND OPTICAL 
    DOUBLE HOT SPOT OF 3C 351}


\author{G.Brunetti\altaffilmark{1,2}, M.Bondi\altaffilmark{1},
A.Comastri\altaffilmark{3}, M.Pedani\altaffilmark{4},
S.Varano\altaffilmark{2}, 
%R.Fanti\altaffilmark{1,5}, 
G.Setti\altaffilmark{1,2}, M.J.Hardcastle\altaffilmark{5}}


\altaffiltext{1}{Istituto di Radioastronomia del CNR, via Gobetti 101,
I--40129 Bologna, Italy}
\altaffiltext{2}{Dipartimento di Astronomia, Universita' di
Bologna, via Ranzani 1, I--40127 Bologna, Italy}
\altaffiltext{3}{Osservatorio Astronomico di Bologna, via Ranzani 1,
I--40127 Bologna, Italy}
\altaffiltext{4}{Centro Galileo Galilei - S/C La Palma, 38700 TF Spain}
\altaffiltext{5}{Department of Physics, University of Bristol,
Tyndall Avenue, Bristol BS8 1TL, UK}

\begin{abstract}
In this letter we report a {\it Chandra} X-ray detection of the 
double northern hot spot of the radio quasar 3C 351.
The hot spot has also been observed in the  
optical with the {\it Hubble
Space Telescope} ($R$-band) and with the 3.5m. 
Telescopio Nazionale Galileo ($B$-band).
The radio-to-optical and X-ray spectra  
are interpreted as the results of the synchrotron and 
synchrotron-self-Compton (SSC)
mechanisms, respectively, 
with hot-spot magnetic field strengths $\sim$3 times smaller
than the equipartition values. 
In the framework of shock acceleration theory, we show 
that the requirement for such a relatively small field strength is in
agreement with the fitted synchrotron spectral models and
with the sizes of the hot spots.
Finally, we show that the combination of a lower magnetic field 
strength with the high frequencies of the synchrotron cut-off 
in the fitted synchrotron spectra provides 
strong evidence for electron acceleration in the hot spots.   
\end{abstract}

\keywords{quasars: individual (3C351) --- acceleration of particles ---
radiation mechanisms: non--thermal}

\section{INTRODUCTION}

It is generally assumed that the relativistic particles in 
powerful radio galaxies and quasars, 
initially generated in the vicinity of the active galactic 
nucleus and channeled by the radio jets out to
hundreds of kpc, are re-accelerated in the radio
hot spots, which mark the location of strong planar 
shocks formed at the heads of the jets 
themselves
(e.g., Begelman et al.\ 1984, Meisenheimer et al.\ 1989).
One of the most important pieces of evidence in favor of electron 
re-acceleration in the hot spots is the detection of synchrotron
emission in the optical/near-IR band, due to 
high energy electrons (Lorentz factor $\gamma > 10^5$) 
with very short radiative lifetimes 
(e.g., Meisenheimer et al.\ 1997 and references therein).
X-ray observations of non-thermal emission from the radio hot spots
are of fundamental importance to constrain the energetics and the
spectrum of the relativistic electrons and to test the re-acceleration
scenario.  Until the advent of {\it Chandra} non-thermal X-ray
emission had only been detected from a few hot spots. One well-known
detection was the {\it ROSAT} PSPC observation of the radio hot spots
of Cygnus A (Harris et al.\ 1994), where the hot spot emission was
interpreted as synchrotron-self-Compton (SSC) emission, implying
magnetic field strengths in the hot spots close to the equipartition
values.  {\it Chandra} has enabled significant progress in this field,
with a number of successful detections in the first 2 years of
observations (3C 295: Harris et al.\ 2000; Cyg A: Wilson et al.\ 2000;
Pictor A: Wilson et al.\ 2001; 3C 123: Hardcastle et al.\ 2001; 3C
207: Brunetti et al.\ 2001; 3C\,263, Hardcastle et al.\ in
prep.). With the exception of Pictor A, the best and most
straightforward interpretation for the X-ray emission in these
objects is provided by the SSC mechanism under approximate
equipartition conditions.

\noindent
In this letter we report on the {\it Chandra}
discovery of X-ray emission from the double northern 
hot spot of 3C 351 (components J and L of Bridle et al. 1994)
and on the modeling of the broad-band
spectrum.
$H_0=50$ km s$^{-1}$ Mpc$^{-1}$ and
$q_0=0.5$ are assumed throughout; 1 arcsec corresponds
to 6.2 kpc at the redshift of 3C 351.
Reported errors are at the 90\% confidence level.

\section{TARGET AND DATA ANALYSIS}

The powerful double-lobed
radio source 3C 351 is identified with a 
quasar at a redshift $z=0.371$ (Laing et al.\ 1983).
In addition to the {\it Chandra}
data, we have analyzed Very Large Array (VLA) and {\it HST} archive data 
and obtained
new $B$-band observations with the 3.5m Telescopio Nazionale Galileo
(TNG).
%, with the aim of comparing the radio, optical and X-ray 
%spectra of the hot spots.

\subsection{Radio and optical data}

Table 1 lists the observations used in this paper and the derived
flux densities at different frequencies. 
The 1.4 GHz values are derived from an A+B-configuration VLA 
image published by
Leahy \& Perley (1991).
Excellent high-resolution deep radio images at 5 GHz have been published
by Bridle et al.\ (1994), and the 5-GHz flux density
of 3C 351--J is from this paper.
Higher frequency radio fluxes have been obtained by reducing
archive VLA observations at 15 and 22 GHz.
Standard procedures for calibration and deconvolution were applied 
to the data. The fluxes of the hot spots were corrected for primary beam
attenuation and for atmospheric opacity.

\noindent
R\"oser (1989) first discovered the northern radio hot 
spots in the optical band, whereas L\"ahteenm\"aki \& Valtaoja (1999) 
found optical linear polarisation ($I$- and $V$-band) 
with the 2.5 m Nordic Optical Telescope on La Palma; 
to our knowledge, no optical fluxes have been 
quoted in the literature.
On 1995 November 29, 3C 351 was observed with the WFPC2 on the {\it HST}
for 2400 s using the filter F702W (close to $R$-band).
We analyzed these {\it HST} observations in the standard manner,
removing cosmic rays and using the bright quasar nucleus to correct the
astrometry.
On the basis of positional coincidence we identified optical
counterparts of the two northern radio hot spots on the WFPC2 image.
In order to better constrain the optical spectra we also observed
these hot spots with a 1200-s B-band exposure at 
the 3.5m TNG on La Palma during the night of 2001 August 16.
The hot spots were detected with a high signal to noise and the
errors on the fluxes (Table 1), obtained with 4-arcsec aperture 
photometry (the seeing was 1.5 arcsec), are within 10\%.

\subsection{X-ray data}

We have analyzed archival data on 3C 351
observed for 9.8 ks
with the {\it Chandra} observatory on 2000 June 1
in GTO time.
The raw level 1 data were
re-processed using the latest version (CIAO2.1) of the
CXCDS software.
%The target was near the aim point on the S3 ACIS chip and
%thus the on-axis point spread function is applicable.
We generated a clean data set by selecting the standard grades
(0,2,3,4,6) and energies in the band 0.1--10 keV.
The X-ray image is shown in Fig. 1, with superimposed contours of 
the 1.4-GHz VLA radio image.
In addition to the nuclear source,
two relatively bright X-ray sources, coincident
with the double northern radio hot spot,
are clearly detected: $\sim$ 82$\pm$10 and 54$\pm$9 net counts
are associated
with components J and L, respectively.
Despite the poor statistics, we have attempted to derive
spectral information using appropriate response and
effective area functions. In both cases a single
power law (3C 351--J: $\alpha=1.2\pm0.5$;
3C 351--L: $\alpha=1.7\pm1.2$), with absorption fixed at the Galactic
value $N_{\rm H}=2.26\times 10^{20}$cm$^{-2}$ (Elvis et al.\ 1989),
provides an acceptable description of the data.

\section{EMISSION PROCESSES AND ELECTRON ACCELERATION}

\subsection{Modeling the broad band spectrum}

As no bright thermal emission from a cluster
around 3C 351 has been detected by the {\it Chandra} observation, 
we find that X-ray emission from any shocked gas in the hot spot
region would be orders of magnitude lower than that observed.
As discussed in the Introduction, the most plausible non-thermal  
mechanism responsible for X-ray emission is SSC.
In Fig. 2 we show fits to the broad band 
spectrum of the hot spots 3C351--J and L.
The emitted synchrotron and SSC spectra 
are obtained with standard 
recipes, assuming an electron energy distribution as
derived under basic Fermi-I acceleration theory: 
a power-law spectrum with injection energy index $\delta$
which steepens at higher energies ($\gamma>\gamma_{\rm b}$) 
up to a high 
energy cutoff $\gamma_{\rm c}$, whereas it 
flattens at lower energies before a low energy cutoff 
(equations for the electron spectrum and for the SSC calculation 
are given in Brunetti et al. 2001, 
Eqs. 2--9 and Appendix A).
The SSC flux is obtained by taking radii 
$R_{\rm J}=0.1$ and $R_{\rm L}=0.5$ arcsec for 
3C 351--J and L, respectively
(Bridle et al.\ 1994).
The parameters of the spectra obtained from
fitting the radio and optical data points, i.e.,
the electron injection index ($\delta$), the
synchrotron break ($\nu_{\rm b}$) and
cutoff frequency ($\nu_{\rm c}$)
are given in Table 2 (90\% confidence level).
%The reported ranges for $\nu_{\rm b}$ and $\nu_{\rm c}$
%are calculated at the 90\% confidence level.
A synchrotron spectrum with a cutoff frequency at very high
energies is still consistent with the data of 3C 351--J
and is also plotted in Fig.\ 2 in order to show the maximum synchrotron
contribution to the X-rays ($\leq 30\%$ at 90\% confidence level).
It should be noted that the high-frequency
cutoff of 3C 351--J 
is at least one order of magnitude
larger than those of other optical hot spots in the
literature, and that the break frequencies 
for both hot spots 
are also higher than the typical breaks measured 
in the synchrotron spectra
of optical hot spots (e.g., Meisenheimer et al.\ 1997; 
Gopal-Krishna et al.\ 2001).

\noindent
We find that for both hot spots
the SSC mechanism provides a good representation of the
{\it Chandra} data for magnetic fields, ($B_{\rm ic}$, Table 2),
which are a factor of $\sim 3$
smaller than the equipartition values ($B_{\rm eq} \simeq$230
$\mu$G for 3C 351--J and $B_{\rm eq} \simeq$100 $\mu$G for
3C 351--L).
For these field strengths, we find that inverse-Compton (IC)
scattering of cosmic microwave background (CMB) photons
could account for $\sim 10$\% of the X-ray flux of
3C 351--L, whereas it does not significantly
contribute to the X-ray spectrum of 3C 351--J.
It is interesting to note that the X--ray luminosity of
3C 351--J is comparable to or slightly higher than the
synchrotron radio to optical luminosity.
If the X-rays are of SSC origin, 
the radiative losses of the relativistic 
electrons would be dominated by IC 
(due to second- and third-order scattering) so that 
the electrons' cooling times would be a factor of 
$\sim 4$ shorter than those due to the synchrotron 
process alone.
However, as the value of the magnetic 
field strength in the Compton-dominated
3C 351--J is $\sim 3$ times smaller than equipartition, 
the radiative 
cooling time of the electrons emitting synchrotron 
radiation at $\sim 10^{16}$Hz is a factor $\sim 2.5$ larger  
than that computed under equipartition, so that
the acceleration of such particles is eased with respect to the 
equipartition case.
A viable model for the X--ray emission of 3C 351--J can also
be obtained with a combination of a synchrotron spectrum and a
SSC component (accounting for $\geq70\%$ of the flux).
In this case, the resulting spectral index would be somewhat
steeper than that of the SSC model. 
This would imply very efficient electron acceleration processes,
but the basic SSC model remains unchanged.
Improved spectral measurements may be able
to settle the size of a possible synchrotron
contribution.
%
%\subsection{The role of boosting}

Relativistic boosting appears to be necessary
to explain the X-ray jet emission in
a few core-dominated radio-loud quasars
(e.g., Tavecchio et al.\ 2000; Celotti et al.\ 2001;
Brunetti et al.\ 2001), however its role
%the importance of boosting 
in the case of the hot spots
%of radio sources 
is still unclear; for instance, statistical analysis of
samples of FRII radio sources implies non-relativistic velocities
(e.g., Arshakian \& Longair 2000).
Under equipartition (in the hot spot frame) 
IC scattering of CMB photons can produce the observed
X-ray fluxes for Doppler factors ${\cal D} \sim 8$ 
and 3.8 in the cases of 3C 351--J and L, respectively.
Such factors, however, can only be obtained for
relatively small angles between the hot spots' velocities 
and the line of sight ($\theta \leq 6^o$ with $\Gamma_{\rm bulk} > 4$
for 3C 351--J, and 
$\theta \leq 15^o$ with $\Gamma_{\rm bulk} > 2$
for 3C 351--L), whereas the radio properties of 3C 351 and
the lack of lobe asymmetry suggest much
larger inclination angles and smaller hot spot
velocities.
This problem remains even if we relax the equipartition 
condition: assuming a departure 
from equipartition
similar to that required by the SSC, 
a Doppler factor ${\cal D} \sim 4.5$
($\theta \leq 10^o$) is necessary to match the X-ray flux 
of 3C 351--J by IC scattering of the CMB.
A second possible way of accounting for the large X-ray fluxes
of the hot spots is to require the SSC to be enhanced with respect to the 
synchrotron emission via Doppler de-boosting.
We find, however, that in order to match the data under 
equipartition conditions 
${\cal D} \sim 0.25$ is required, implying extremely 
high bulk Lorentz factors ($\Gamma_{\rm bulk} \sim 15-30$,  
with $\theta  \sim 30-40^\circ$).
%and 
%a very high magnetic field strength ($B \sim 900 \mu$G)
%in the case of 3C 351--J. 
 
\subsection{Is {\it in situ} re-acceleration necessary ?}

Based on evidence for relativistic jet bulk motion
out to 100-kpc scales, 
Gopal-Krishna et al.\ (2001) have reconsidered 
a minimum loss scenario in which the relativistic electrons,  
accelerated in the central active nucleus,
flow along the jets losing energy only 
due to the inescapable IC scattering of CMB photons.
Under these assumptions, comparing the electron radiative lifetime 
with the travel time to the
hot spots, they find that {\it in situ}
electron re-acceleration is in general
not absolutely necessary to explain the optical/near-IR synchrotron
radiation from the hot spots so far detected.
Following Gopal-Krishna et al.\ (2001) we define $\eta=
D_{\rm obs}/D_{\rm max}$, the ratio between the 
distance of the hot spots from the nucleus and the  
largest distance covered by the electrons in the jet 
before dropping to less than e$^{-1}$ of their initial energy
due to IC losses; i.e., 

\begin{equation}
D_{\rm max}({\rm kpc}) \simeq 
145 {{ \beta_{\rm bulk} \sin \theta }\over
{\Gamma_{\rm bulk} (1+z)^{4.5} }}
\left[ {{ B(\mu G) }\over{\nu_{\rm c}(10^{14}{\rm Hz}) }}
\right]^{1/2}
\label{dmax}
\end{equation}

\noindent
With $\theta \sim 30^\circ$ and our constraints on $\nu_{\rm c}$
and $B$ we find $\eta\geq 26$--$110$ (3C 351--J) and
$\eta\sim 9$--$38$ (3C 351--L) for $\Gamma_{\rm bulk}=2$--$10$
(as adopted by Gopal-Krishna et al. 2000).
We also find that if the electrons are accelerated in 3C 351--J and then
transported in a jet out to
3C 351--L, the lifetime of the optically emitting
electrons in 3C 351--L ($\sim 5\times 10^4$ yr assuming
$B \geq 10 \mu$G in the jet) is at least one order of magnitude
less than the time necessary to cover the distance between the two
hot spots at non-relativistic speeds, and $\eta$ is still $>2.2$ 
if they are transported at relativistic speeds.
These large values unequivocally point to the necessity of
efficient re-acceleration processes in both the northern 
hot spots of 3C 351.

\subsection{Electron acceleration in the hot spot}

An independent
estimate of the magnetic field strength $B$ in the hot spot
can be obtained in the framework of shock acceleration models.
The highest energies of the relativistic electrons accelerated
in a shock region by Fermi processes is given by the ratio
between gain and loss terms.
For a strong shock (here, for simplicity, we consider 
the non-relativistic case) 
the largest energy of the accelerated electrons is
(e.g., Meisenheimer et al.\ 1989):

\begin{equation}
\gamma_{\rm c}
\propto
u_{-}^2 B_+^{-2} \lambda_{+}(B_+)^{-1}
\label{gammac}
\end{equation}

\noindent
where $u_-$ is the velocity of the flow in the
upstream region (with respect to the shock), 
$B_+=\sqrt{ B_{\rm ic}^2 + B^2}$ 
is the total equivalent magnetic field strength 
in the downstream region, 
and $\lambda_+(B_+)$ is the mean free path of
the relativistic electrons in the downstream region.

\noindent
The break energy of the electrons in the downstream 
region (i.e., the highest energy of the oldest electrons
in the post shock region) is:

\begin{equation}
\gamma_{\rm b}=
{{ \gamma_{\rm c} }\over{ 1 + C {{ l }\over{u_+}}
B_+^2 \gamma_{\rm c} }} 
\propto {{ u_+ }\over{l}} B_+^{-2}
\label{gammab}
\end{equation}

\noindent
where $u_+$ is the velocity of the flow in the
downstream region, $l$ is the linear extension 
of such a region and $C$ a constant.
Substituting into Eqs.(\ref{gammac}--\ref{gammab})
the expression for the synchrotron
cutoff frequency, $\nu_{\rm c} \propto \gamma_{\rm c}^2
B$, one obtains:
%an expression for $B_+$ :

\begin{equation}
B_+
\sim 2.3
\left( {{ u_+ }\over{0.3}} 
[ {{ \gamma_{\rm c} }\over{
\gamma_{\rm b} }} -1 ] \right)^{2/3}
l_{\rm kpc}^{-2/3} \left( {{ \nu_{\rm c} }\over{
10^{15}{\rm Hz} }} \right)^{-1/3}
\xi^{1/3}
\label{bp}
\end{equation}

\noindent
where $\xi = B/B_+$ is $\simeq$ 0.5 for
the hot spot J and 0.9 for L, and
the value of $l_{\rm kpc}$ can be estimated from
Bridle et al. (1994).
For $\gamma_{\rm c}/\gamma_{\rm b} >> 1$ Eq.4 reduces to
$B_+ \propto \nu_{\rm b}^{-1/3}$, and
from the values of $\nu_{\rm b}$
previously constrained we find
$B \sim 80$ and $30 \mu$G for J and
L, respectively, with a formal uncertainty
of about a factor 2.
These values are in good
agreement with the magnetic field strengths implied by an SSC origin
for the X-ray emission.

\section{CONCLUSIONS}

The radio to optical spectrum of the two northern hot spots
of 3C 351 is well accounted
for by a synchrotron model under the 
standard shock acceleration scenario.
The most straightforward interpretation of the X-ray emission discovered
by {\it Chandra} is the SSC mechanism.
A viable model for 3C 351--J is also given by a combination of SSC and
synchrotron spectra, with the SSC accounting for $\geq 70\%$
of the X-ray flux.
In order to match the X-ray fluxes a magnetic field strength 
of a factor 3 lower than the equipartition value is required in 
both the hot spots.
In principle the assumption of relativistic bulk motion of the hot
spots might reduce the departure from equipartition, but large {\it ad
hoc} distortions of the radio jet would be required, with the hot
spots moving close to the line of sight.
The frequencies of the cutoff
in the synchrotron spectra and the  
magnetic field strengths 
point to the need for {\it in situ} electron 
(re)acceleration in the region of both the hot spots.
When we independently estimate the magnetic field strength 
of the hot spots, using relationships from standard shock theory,
our results are consistent with the more precise 
value obtained by modeling the X-ray
emission as resulting from the SSC mechanism.

\acknowledgments

This work is partially supported by the Italian Ministry
for University and Research (MURST) under grant Cofin99-02-37
and Cofin00-02-36.
This paper is 
based on observations made with the Italian Telescopio Nazionale
Galileo (TNG) operated on the island of La Palma by the 
Centro Galileo Galilei of the CNAA 
at the Spanish Observatorio del Roque de los Muchachos of 
the Instituto de Astrofisica de Canarias.
We would like to thank R.\ Fanti for very useful discussions and
A.\ Zacchei for technical support.

\begin{table}
\begin{center}
\caption{Observations}
\begin{tabular}{cccccccc}
\tableline\tableline
Band & Freq & Telescope & Flux J & Flux L \\
     &      &           & mJy    & mJy    \\
\tableline
      Radio & 1.4 & VLA A+B & 480$^a$  & 1290$^a$ \\
             & 4.9 & VLA A+B & 201$^b$  &        \\
	            &15.0 & VLA C   &  85  &  219     \\
		           &22.0 & VLA C   &  72  &  174     \\
\tableline
			   Optical& R & HST    &2.5E-3&3.8E-3       \\
			          & B & TNG    &1.5E-3&1.3E-3       \\
\tableline
	 X-ray & 1 keV& {\it Chandra}&5.2$\pm$1.1E-6&3.4$\pm$1.1E-6 \\
\tableline
\end{tabular}
\end{center}
$^{a}$ Leahy \& Perley (1991), $^{b}$ Bridle et al.\ (1994).
Errors on radio fluxes (1.4 to 15 GHz) are within 5\%, whereas those
on 22 GHz and optical fluxes are within 10\%.
\end{table}

\begin{table}
\begin{center}
\caption{Spectral parameters and $B$-fields}
\begin{tabular}{cccccccccc}
\tableline\tableline
Hot Spot & $\delta$ & $\nu_{\rm b}$ & $\nu_{\rm c}$ & B$_{\rm ic}$ \\
         &          &  ($10^{13}$Hz)&($10^{14}$Hz) &   ($\mu$G) \\
\tableline
  L    & 2.5 & 3.6$\pm$3.0  &  11$\pm$7 & 35 \\
  J    & 2.4 & 0.35$\pm$0.25  & $\geq$90 & 70 \\
\tableline
\end{tabular}
\end{center}
\end{table}

\newpage

\begin{figure}
\plotone{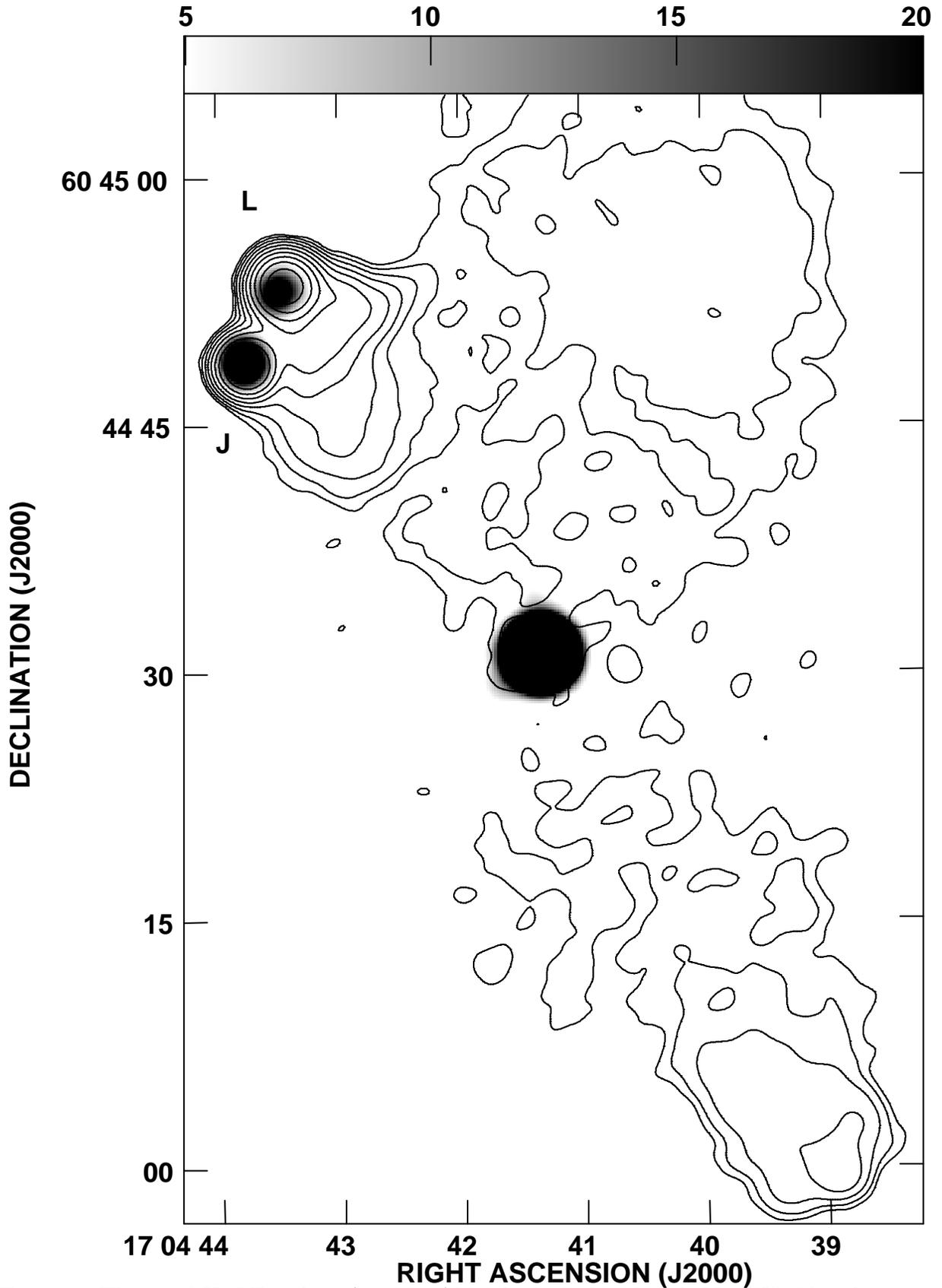}
\caption{The 1.4 GHz VLA data (contours) overlaid on the
0.3--8 keV ACIS {\it Chandra} image (grey scale).
The radio contours are : 1.0 $\times$ (-1, 1, 2, 4,{\ldots} )
mJy/beam; beam=1.85 arcsec.}
\end{figure}

\begin{figure}
\plotone{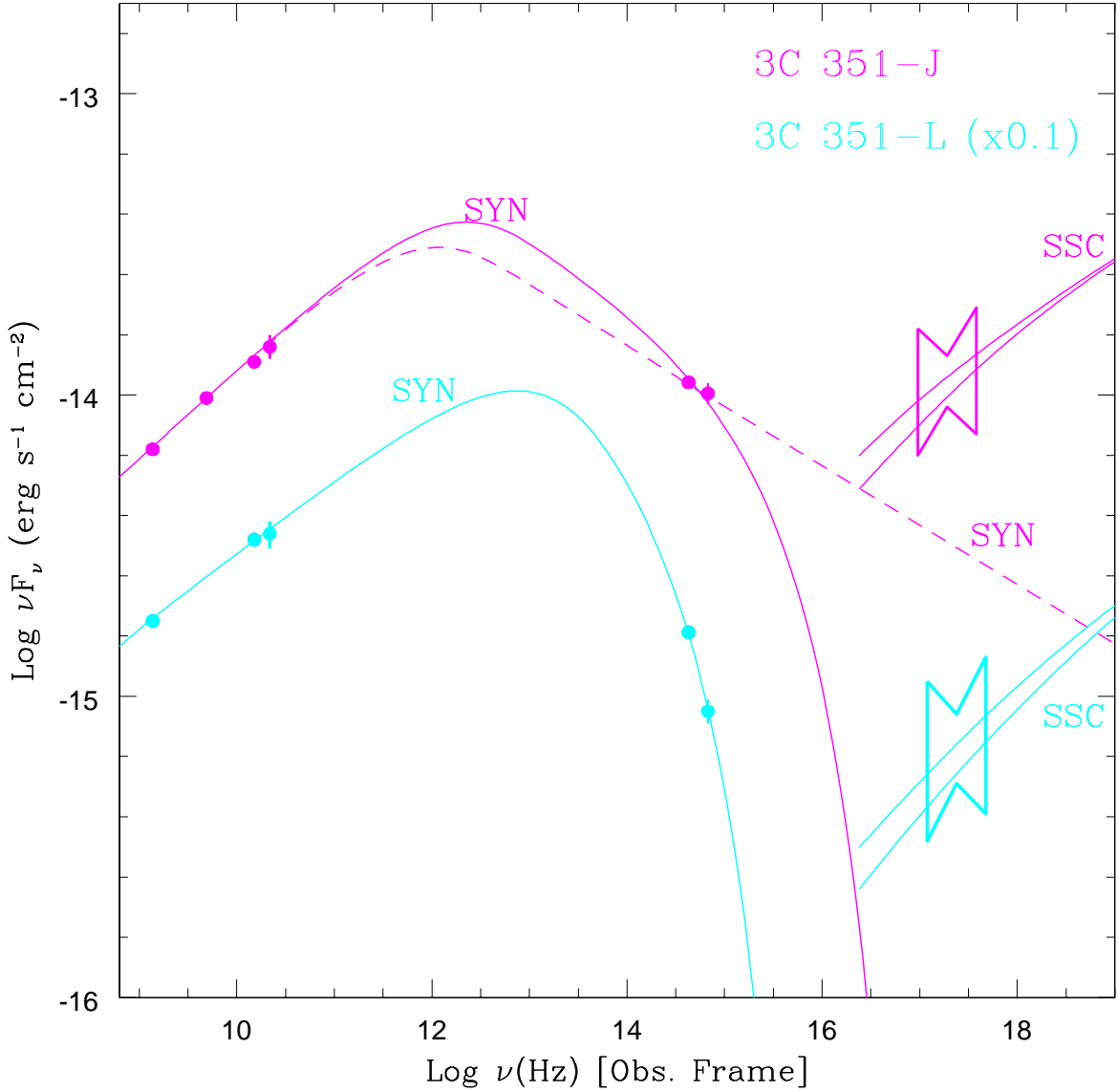}
\caption{Synchrotron and SSC models (solid lines) compared with
the data of 3C 351--J (magenta) and L (cyan).
3C 351--J: the synchrotron model (solid line) 
has $\delta=2.4$,
$\nu_{\rm b}=3.2\times 10^{12}$Hz and $\nu_{\rm c}=1.2\times 10^{16}$Hz.
The corresponding SSC models are calculated with $B=70 \mu$G
and with a low frequency cutoff in the synchrotron photons at
100 MHz (lower curve) and 1 MHz (upper curve).
A synchrotron model with $\nu_{\rm b}=1.6\times 10^{12}$Hz and
no cutoff is also shown (dashed line).
3C 351--L: the synchrotron model has
$\delta=2.5$,
$\nu_{\rm b}=2.2\times 10^{13}$Hz and $\nu_{\rm c}=1.1\times 10^{15}$Hz,
the corresponding SSC models are calculated with $B=35 \mu$G,
with the other parameters as in the case of 3C 351--J.
The fluxes of 3C 351--L are reported in the panel
multiplied by a factor 0.1.}
\end{figure}

\end{document}